\documentclass[aps,superscriptaddress,prd,twocolumn,nofootinbib]{revtex4}

\usepackage{amsmath,bm}
\usepackage{graphicx}
\usepackage{epstopdf}
\usepackage{subfigure}
\usepackage{epsfig}
\usepackage{amsmath,amssymb,amsfonts}
\usepackage{multirow}
\usepackage{tabularx}
\usepackage{color}
\usepackage{palatino}
\usepackage[breaklinks,colorlinks,urlcolor=blue,citecolor=blue,linkcolor=blue]{hyperref}
\usepackage[utf8]{inputenc}
\usepackage{verbatim}
\usepackage{array}
\usepackage{marvosym}
\usepackage{float}
\def\bea#1\eea{\begin{align}#1\end{align}} 
\newcommand{\nnu}{\nonumber\\}
\newcommand{\bef}{\begin{figure}[htb]\centering}
\newcommand{\eef}{\end{figure}}
\begin{document}
\title{Jet radius and momentum splitting fraction with \\ dynamical grooming in heavy-ion collisions}

\date{\today  \hspace{1ex}}

\begin{abstract}
We investigate the medium modifications of momentum splitting fraction and groomed jet radius with both dynamical grooming and soft drop algorithms in heavy-ion collisions. In the calculation, the partonic spectrum of initial hard scattering in p+p collisions is provided by the event generator PYTHIA 8, and the energy loss of fast parton traversing in a hot/dense QCD medium is simulated with the Linear Boltzmann Transport (LBT) model. 
We predict the normalized distributions of the groomed jet radius $\theta_g$ and momentum splitting fraction $z_g$ with the dynamical grooming algorithm in Pb+Pb collisions at $\sqrt{s_{\mathrm{NN}}}$ = 5.02 TeV,
then compare these quantities in dynamical grooming at $a=0.1$, with that in soft drop at $z_{\mathrm{cut}} = 0.1$ and $\beta = 0$. It is found that the normalized distribution ratios Pb+Pb/p+p with respect to $z_g$ in $z_{\mathrm{cut}} = 0.1$, $\beta = 0$ soft drop case are close to unity and those in  $a=0.1$ dynamical grooming case show enhancement at small $z_g$, and Pb+Pb/p+p with respect to $\theta_g$ in the dynamical grooming case demonstrate weaker modification than those in the soft drop counterparts. 
We further calculate the groomed jet number averaged momentum splitting fraction $\rm \langle z_g \rangle_{jets}$ and averaged groomed jet radius $\rm \langle \theta_g \rangle_{jets}$ in p+p and A+A for both grooming cases in three $p^{\rm ch, jet}_{\rm T}$ intervals, and find that the originally generated well balanced groomed jets will become more momentum imbalanced and less jet size narrowing due to jet quenching, and weaker medium modification of $z_g$ and $\theta_g$ in $a =0.1$ dynamical grooming case than in the soft drop counterparts. 
\end{abstract}

\author{Lei Wang}
\affiliation{Key Laboratory of Quark and Lepton Physics (MOE) and Institute of Particle Physics, Central China Normal University, Wuhan 430079, China}
\author{Jin-Wen Kang}
\affiliation{Key Laboratory of Quark and Lepton Physics (MOE) and Institute of Particle Physics, Central China Normal University, Wuhan 430079, China}
\author{Qing Zhang}
\affiliation{Key Laboratory of Quark and Lepton Physics (MOE) and Institute of Particle Physics, Central China Normal University, Wuhan 430079, China}
\author{Shuwan Shen}
\affiliation{Key Laboratory of Quark and Lepton Physics (MOE) and Institute of Particle Physics, Central China Normal University, Wuhan 430079, China}
\author{Wei Dai}
\email{weidai@cug.edu.cn}
\affiliation{School of Mathematics and Physics, China University of Geosciences (Wuhan), Wuhan 430074, China}
\author{Ben-Wei Zhang}
\email{bwzhang@mail.ccnu.edu.cn}
\affiliation{Key Laboratory of Quark and Lepton Physics (MOE) and Institute of Particle Physics, Central China Normal University, Wuhan 430079, China}
\affiliation{Guangdong Provincial Key Laboratory of Nuclear Science, Institute of Quantum Matter,South China Normal University, Guangzhou 510006, China}
\author{Enke Wang}
\affiliation{Key Laboratory of Quark and Lepton Physics (MOE) and Institute of Particle Physics, Central China Normal University, Wuhan 430079, China}
\affiliation{Guangdong Provincial Key Laboratory of Nuclear Science, Institute of Quantum Matter,South China Normal University, Guangzhou 510006, China}
\maketitle

\section{introduction}
\label{sec-int}
In high-energy proton-proton and nucleus-nucleus collisions, a series of new jet substructure observables are measured and analyzed from the experiment at the Large Hadron Collider (LHC). The studies of these jet substructures served as powerful tools to probe the fundamental properties of quantum chromodynamics (QCD) and nucleon structure~\cite{Sterman:1977wj,Altheimer:2012mn,Altheimer:2013yza,Marzani:2019hun}. They also provide new opportunities to probe the properties of the hot dense medium, quark-gluon plasma (QGP) created in heavy-ion collisions~\cite{ALICE:2021njq,Casalderrey-Solana:2019ubu,Vitev:2009rd,Vitev:2008rz,Andrews:2018jcm,ALICE:2019ykw}. The momentum splitting fraction $z_g$ and the jet splitting angle  $\theta_g$ which describe the basic properties of the primary splitting structure of a jet are emerged, and they have been measured in heavy-ion collisions both at RHIC and LHC energies ~\cite{STAR:2021kjt,ATLAS:2019dsv} and some of them are tested against parton energy loss models ~\cite{Wang:2013cia,Majumder:2013re,Zapp:2008gi,Zhang:2003wk,Dai:2018mhw,Dai:2022sjk,Wang:2019xey,Schenke:2009gb,Armesto:2009fj}.

The study of these jet substructures require the definition of the jet grooming techniques~\cite{Dasgupta:2013via, Dasgupta:2013ihk} which is utilized to isolate branches of a jet that correspond to a hard splitting by removing soft wide-angle radiation. Soft drop grooming algorithm~\cite{Larkoski:2014wba,ALargeIonColliderExperiment:2021mqf,Tripathee:2017ybi}, as one of the techniques, is defined by setting a condition $z_g > z_{\mathrm{cut}} \theta_g^\beta$ to remove soft wide-angle radiation inside a jet, thus, the groomed momentum splitting fraction of the two branches $z_g$ and the angle of the remaining branches $\theta_g$ have been studied in experiment~\cite{CMS:2016jys,Kauder:2017cvz} and theory~\cite{Chang:2017gkt,Chien:2016led,Kang:2019prh,Ringer:2019rfk}. Since the soft drop algorithm exhibits the sensitivities to the values of the soft threshold $z_{\mathrm{cut}}$ and the angular exponent $\beta$,
a new method called dynamical grooming~\cite{Mehtar-Tani:2019rrk,Caucal:2021cfb} is proposed to suppress these sensitivities. It is more closely aligned with the intrinsic properties of a given jet and does not require fine-tuning. The grooming condition in dynamical grooming defines a maximum rather than an explicit cut as in soft drop, thus, every dynamically groomed jet will always return a splitting. However, it is possible that a jet does not contain any splitting even after it satisfies the grooming condition in soft drop. It is important to direct compare the jet substructure observables for inclusive jets with two jet grooming algorithms in p+p collisions. In ALICE experiments~\cite{ALICE:2022hyz} measured the normalized momentum splitting fraction $z_g$ and groomed jet radius $\theta_g$ with the dynamical grooming and the soft drop grooming algorithms in proton-proton collisions at $\sqrt{s} = 5.02$ TeV in $60 < p_{\mathrm{T,jet}} < 80$ GeV respectively. It is meaningful to calculate these two observables in Pb+Pb collisions in both two grooming algorithms, since the soft drop case is already in place~\cite{ALargeIonColliderExperiment:2021mqf}, the prediction in the dynamical grooming case is urgently needed. Furthermore, The ALICE reports the first measurement of these two jet substructure observables in $a=0.1$ dynamical grooming case and $z_{\mathrm{cut}} = 0.1$, $\beta = 0$ soft drop grooming case. It would be also interesting to direct compare the medium modification of the groomed jet substructure observables in these two jet grooming cases. 

Therefore, the paper is organized as follows. Firstly, we will introduce the framework used to calculate the normalized groomed jet radius and momentum splitting fraction with the soft drop and dynamical grooming algorithms in p+p and Pb+Pb collisions. Secondly, we present predictions for the distributions of $z_g$ and $\theta_g$ in central 0-30$\%$ Pb+Pb collisions at $\sqrt{s_{NN}} = 5.02$ TeV for inclusive jets with dynamical grooming using different dynamical grooming parameter $a=0.1,1,2$. Then, We compare the medium modification of jet substructures for the groomed jets selected from $a=0.1$ case in the dynamical grooming and from $z_{\mathrm{cut}}=0.1, \beta =0$ case in soft drop. In the end, we systematically investigate the jet substructure observables $z_g$ and $\theta_g$ for inclusive jets in p+p and Pb+Pb collisions at $\rm \sqrt{s}=5.02$~TeV in both $z_{\mathrm{cut}} = 0.1$, $\beta = 0$ soft drop grooming case and $a=0.1$ dynamical grooming case.


\begin{figure*}\centering
\includegraphics[width = 0.7\linewidth]{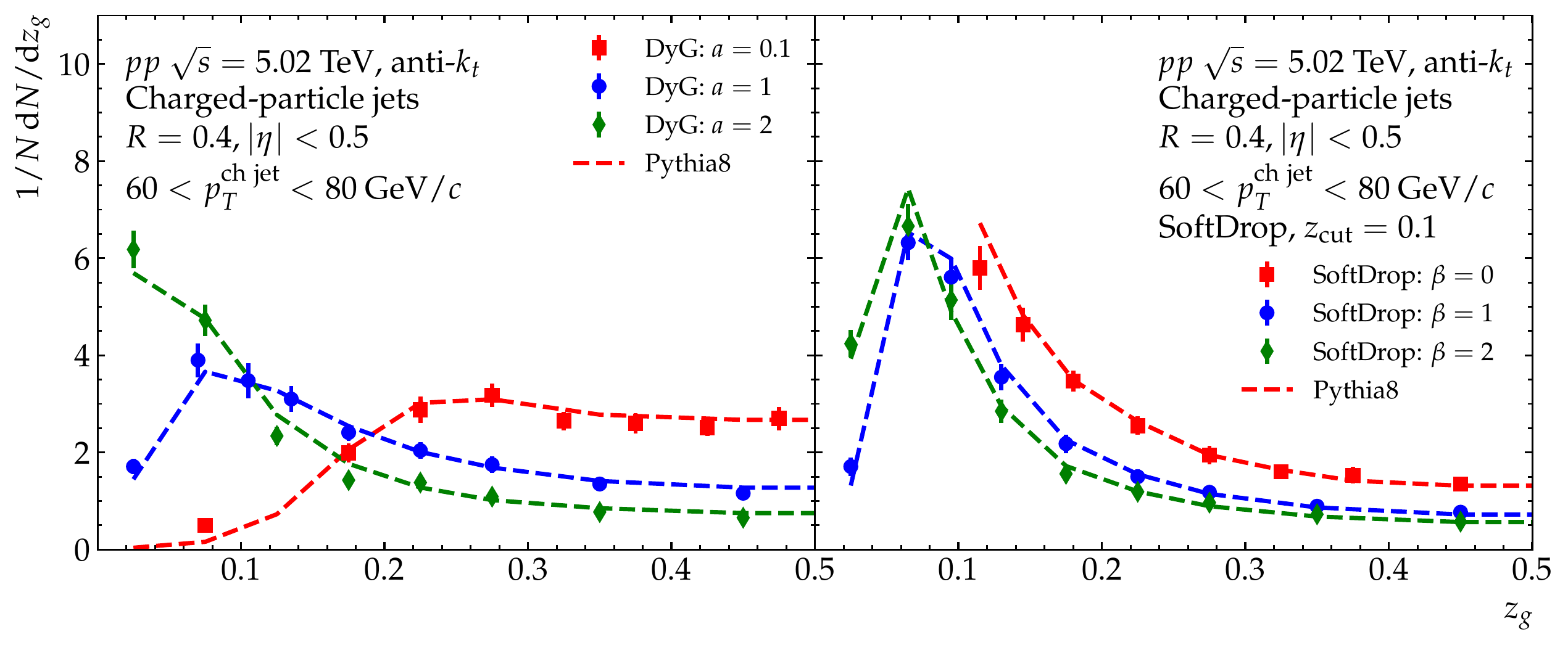}\\
\includegraphics[width = 0.7\linewidth]{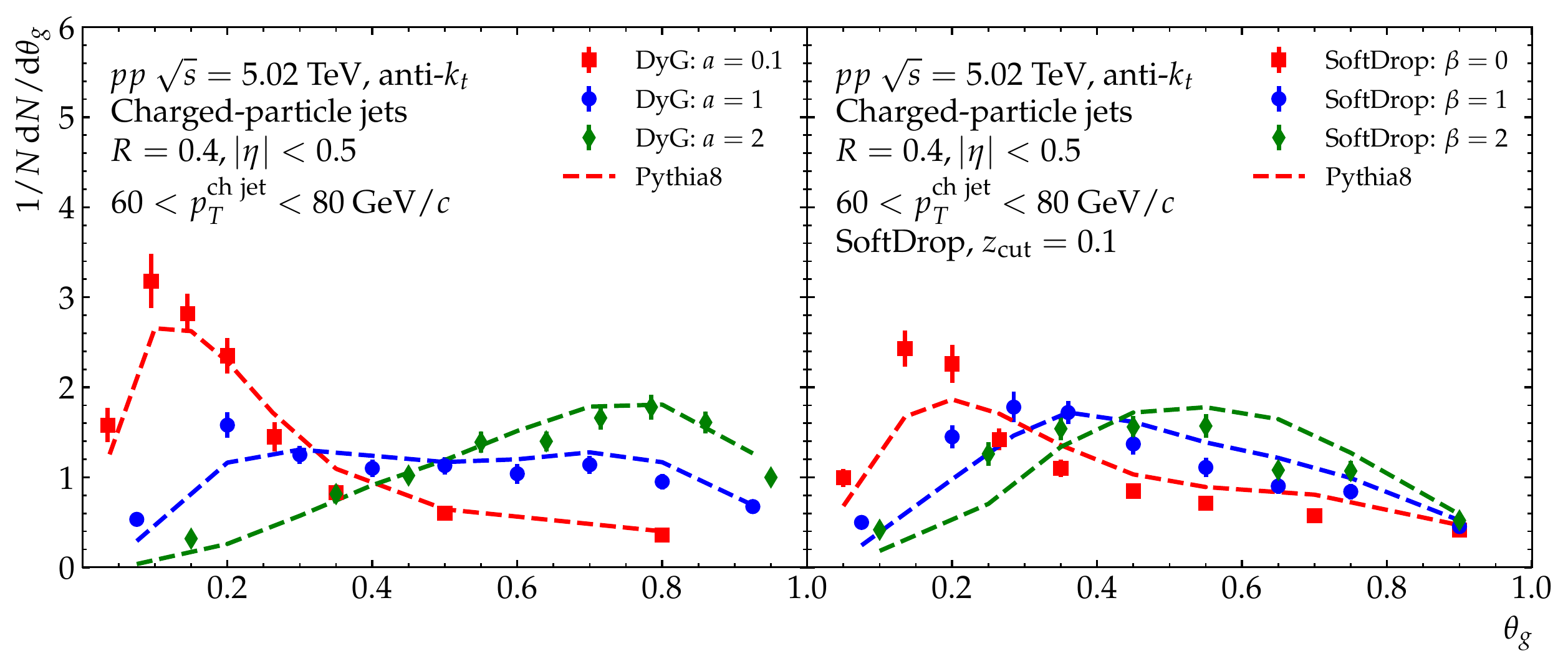}
\caption{Normalized $z_g$ and $\theta_g$ distribution for inclusive jets with the soft drop and the dynamical grooming algorithms in p+p collisions at $\sqrt{s} = 5.02$ TeV from PYTHIA 8 calculation as compared with ALICE data~\cite{ALICE:2022hyz}.}
\label{fig-pp}	
\end{figure*}

\section{analysis framework}

\label{framework}
We start with the impact of the dynamical grooming algorithm and the soft drop grooming algorithm to the jet substructure observables $z_g$ and $\theta_g$ of inclusive jets in p+p collisions at $\sqrt{s} = 5.02$ TeV. The dynamical grooming algorithm identifies a single splitting in the primary Lund plane~\cite{Dreyer:2018nbf} that maximizes 
\bea
z_i(1-z_i)p_{T,i}\theta_i^a ,
\label{1}
\eea
over all splittings in the plane. where 
\bea
z_i &= \frac{p^{\mathrm{sub},i}_{\mathrm{T}}}{p^{\mathrm{leading},i}_{\mathrm{T}}+p^{\mathrm{sub},i}_{\mathrm{T}}} ,
\nnu
\theta_i &= \Delta R_i/R ,
\eea
where $p^{\mathrm{leading},i}_{\mathrm{T}}$ and $p^{\mathrm{sub},i}_{\mathrm{T}}$ are the transverse momenta of the leading jet and subleading jet respectively. $\Delta R_i = \sqrt{\Delta y_i^2+\Delta \phi_i^2}$ is the rapidity-azimuth separation of the splitting, and the exponent $a$ is a free parameter that defines the density of the phase space in the Lund plane that is groomed away. For different parameter $a$, it has different physical significance, the parameter $a = 0$ in the dynamical grooming algorithm corresponds to the case we select the splitting with the most symmetric momentum sharing according to Eq.~\eqref{1}; $a = 1$ refers to the case we prefer the branching with the largest transverse momentum; $a = 2$ means we select the splitting with the shortest formation time~\cite{Mehtar-Tani:2019rrk,Caucal:2021cfb}. Since $a = 0$ is formally collinear unsafe~\cite{Mehtar-Tani:2019rrk}, we use $a=0.1$ instead in this investigation. On the other hand, the soft drop grooming algorithm implements much simpler grooming conditions, $z_i > z_{\mathrm{cut}} \theta_i^\beta$. It is designed to remove soft wide-angle radiation from a jet. The parameter $\beta=0$ in the soft drop grooming algorithm indicates it will groom away splittings below a certain $z_\mathrm{cut}$. 

We computed in Fig.~\ref{fig-pp} the normalized momentum splitting fraction $z_g$ and groomed jet radius $\theta_g$ with the dynamical grooming (left panels) and the soft drop grooming (right panels) algorithms in proton-proton collisions at $\sqrt{s} = 5.02$ TeV in $60 < p_{\mathrm{T,jet}} < 80$ GeV. A Monte Carlo model PYTHIA 8~\cite{Sjostrand:2014zea} is used to generate charged jets. They are reconstructed by the anti-$k_T$ algorithm~\cite{Cacciari:2008gp} with the jet radius $R = 0.4$ and selected according to the same kinematic cuts as adopted by the experimental measurements~\cite{ALICE:2022hyz}. 

Our calculation results are consistent with the experiment data in p+p collisions at $\sqrt{s} = 5.02$ TeV both in the dynamical grooming case and in the soft drop grooming case. In the soft drop grooming case, the $z_g$ with $\beta=0,1,2$ are all tend to distribute at smaller $z_g$, $\beta=0$ case is the most balanced preference among these three. Interestingly, despite of the fact that $a=1,2$ in the dynamical grooming case also tend to the distributed in the smaller $z_g$. $a=0.1$ in the dynamical grooming select the most balanced jets in where the groomed jet samples are much more distributed at larger $z_g$ value. In the right bottom panel in Fig.~\ref{fig-pp}, we find in $\beta = 0$ case which correspond to the most symmetric momentum sharing in the soft drop grooming, the jet sample tend to distribute at the smallest $\theta_g$ region. The larger $\beta$ is, the broader the groomed jet samples are. In the left bottom panel, the most balanced jets selected by dynamical grooming with the parameter $a=0.1$ will be much more concentrated at the smaller $\theta_g$. As the parameter $a$ increases, the radius of the groomed jets tends to be much more wider. Such sensitivity due to the variation of the groomed parameter is much more dramatic than the soft drop case. It is noticed that the case $a=0.1$ selects the splitting with largest $z$, and is similar with $\beta = 0$ case in the soft drop which grooms away splittings below a certain $z$, but $a=0.1$ case in dynamical grooming selects much more balanced groomed jets. It would be interesting to direct compare the medium modification of the groomed jets substructure in these two cases.

\bef
\includegraphics[width = 0.9\linewidth]{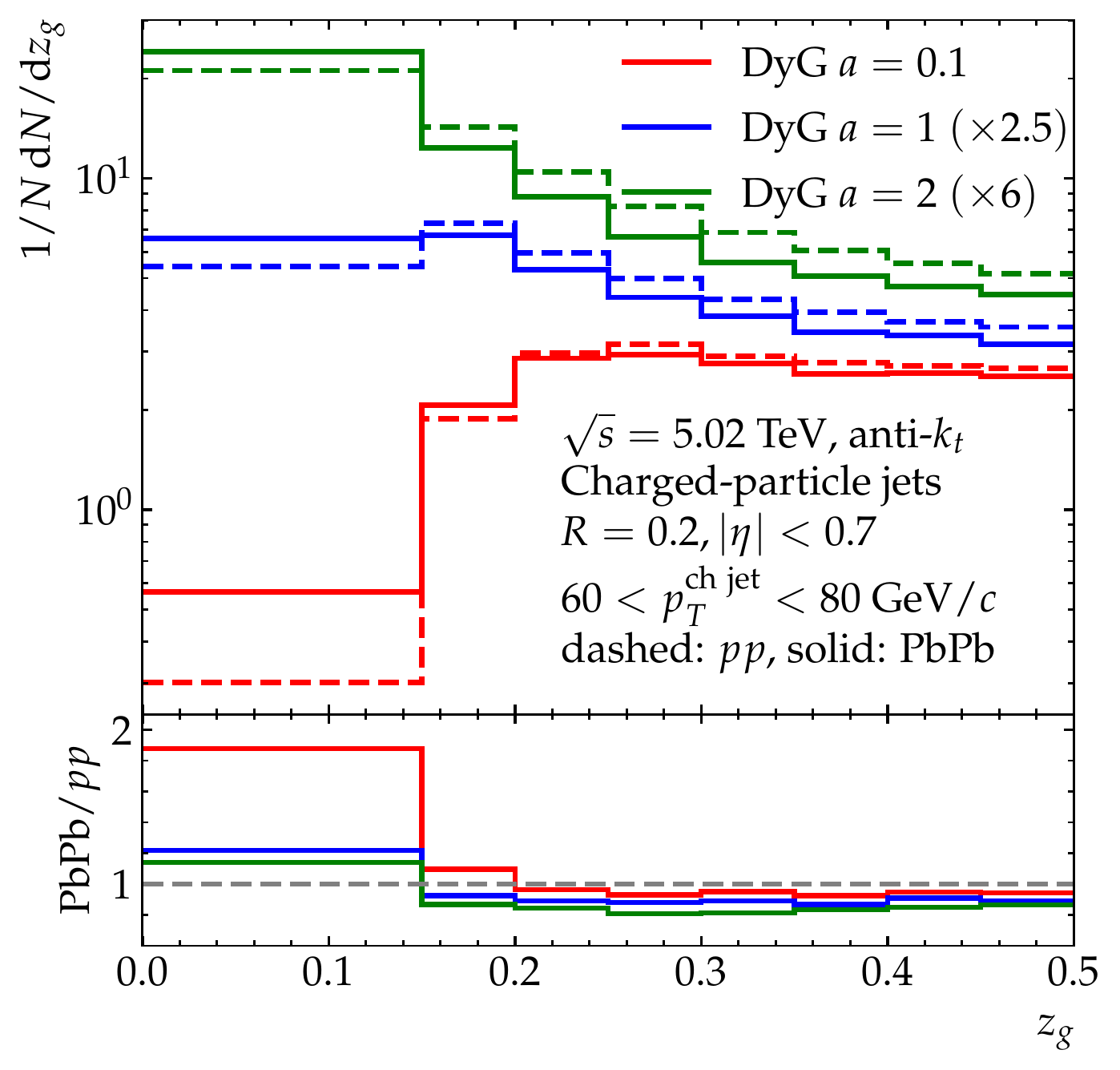}
\includegraphics[width = 0.9\linewidth]{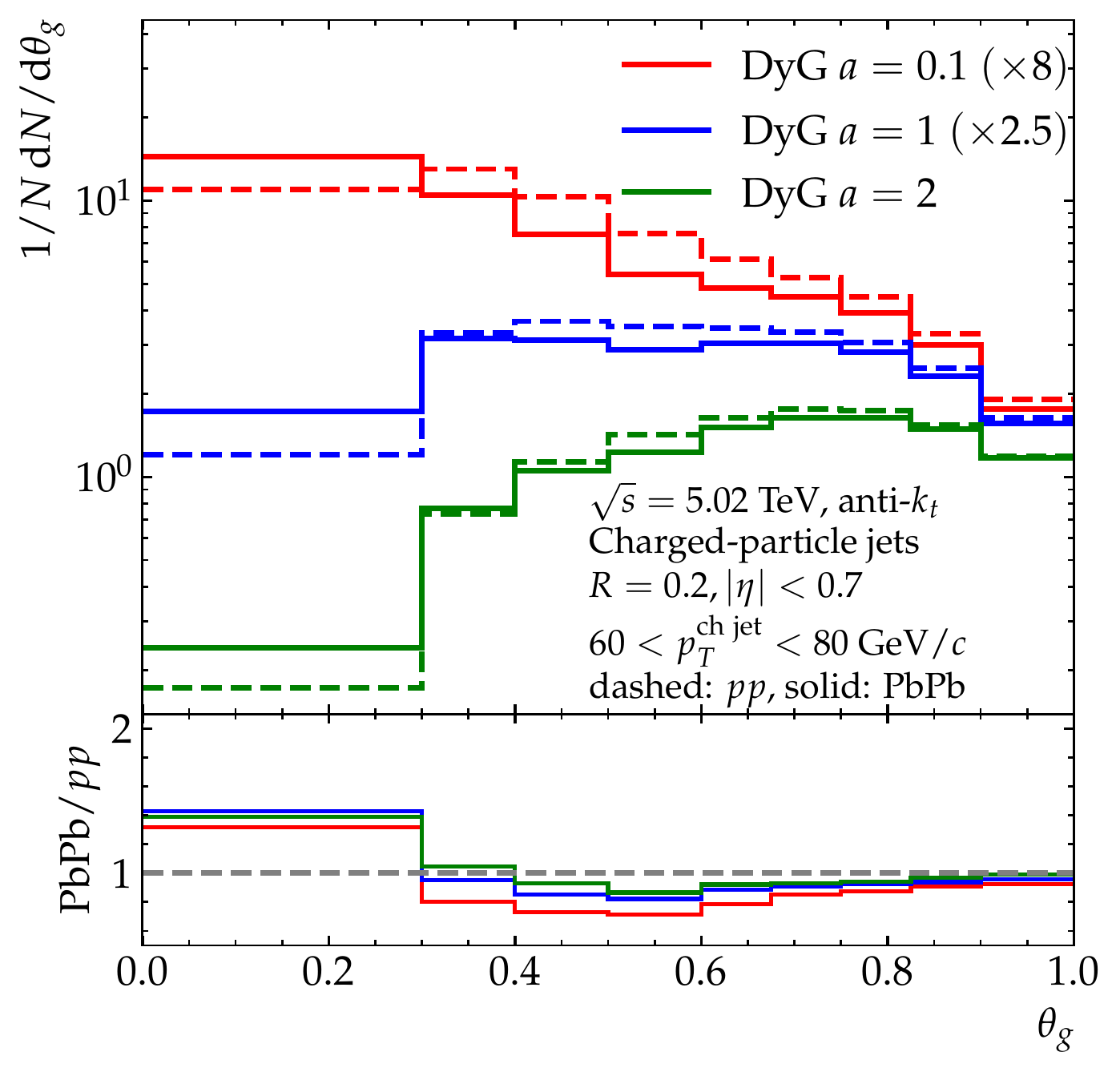}
\caption{Normalized $z_g$ and $\theta_g$ distributions in p+p and $0-30\%$ Pb+Pb collisions at $\sqrt{s} = 5.02$ TeV for inclusive jets with the dynamical grooming algorithm for some values of grooming parameter $a=0.1,1,2$; The dashed line is for p+p collisions and the solid line is for Pb+Pb collisions.}
\label{fig-prediction}	
\eef

The medium modification of jets is simulated within the LBT model ~\cite{Wang:2013cia,He:2015pra,Cao:2016gvr,Cao:2017hhk} that includes both elastic and inelastic processes of parton scattering for jet shower and thermal recoil partons in the QGP medium. The elastic scattering is
described by the linear Boltzmann equation, 
\bea
p_{1} \cdot \partial f_{a}\left(p_{1}\right)=-\int \frac{d^{3} p_{2}}{(2 \pi)^{3} 2 E_{2}} \int \frac{d^{3} p_{3}}{(2 \pi)^{3} 2 E_{3}} \int \frac{d^{3} p_{4}}{(2 \pi)^{3} 2 E_{4}} 
\nnu
\frac{1}{2} \sum_{b(c, d)}\left[f_{a}\left(p_{1}\right) f_{b}\left(p_{2}\right)-f_{c}\left(p_{3}\right) f_{d}\left(p_{4}\right)\right]\left|M_{a b \rightarrow c d}\right|^{2} 
\nnu
\quad \times S_{2}(s, t, u)(2 \pi)^{4} \delta^{4}\left(p_{1}+p_{2}-p_{3}-p_{4}\right)
\eea
where $f_{i}(i=a,b,c,d)$ are the phase-space distributions of medium parton, $S_{2}(s, t, u) = \theta (s\ge 2 \mu_{D}^2) \theta(-s + \mu_{D}^2 \leq t \leq - \mu_{D}^2)$ is a Lorentz invariant condition to regulate the collinear ($t, u \to 0$) divergence of the matrix element $|M_{ab\to cd}|^2$, where $\mu_{D}^2 = g^2 T^2 (N_c + N_f/2)/3$ is the Debye screening mass. The inelastic scattering is described by the higher twist formalism for induced gluon radiation ~\cite{Guo:2000nz,Zhang:2003yn,Zhang:2003wk,Zhang:2004qm}. The  evolution of bulk medium is given by the 3+1D CLVisc hydrodynamical model~\cite{Pang:2012he,Pang:2014ipa} with initial conditions simulated from A Multi-Phase Transport (AMPT) model~\cite{Lin:2004en,Lin:2021mdn}. Parameters used in the CLVisc are fixed by reproducing hadron spectra with experimental measurements. LBT model has been used to provide a nice description of a series of jet quenching measurements, such as bosons tagged jet/hadron production~\cite{Luo:2018pto,Zhang:2021oki}, light and heavy flavor hadrons suppression and single inclusive jets suppression~\cite{He:2015pra,Cao:2016gvr}. After the in-medium evolution, the hadronization of partons is performed by the hadronization method of JETSCAPE~\cite{Putschke:2019yrg} which was based on the Lund srting model~\cite{Sjostrand:1985ys} provided by PYTHIA 8.

\section{numerical results and discussions}

\label{numerical}
\begin{figure*}[hbtp]
\begin{center}
    \includegraphics[width=0.9\textwidth]{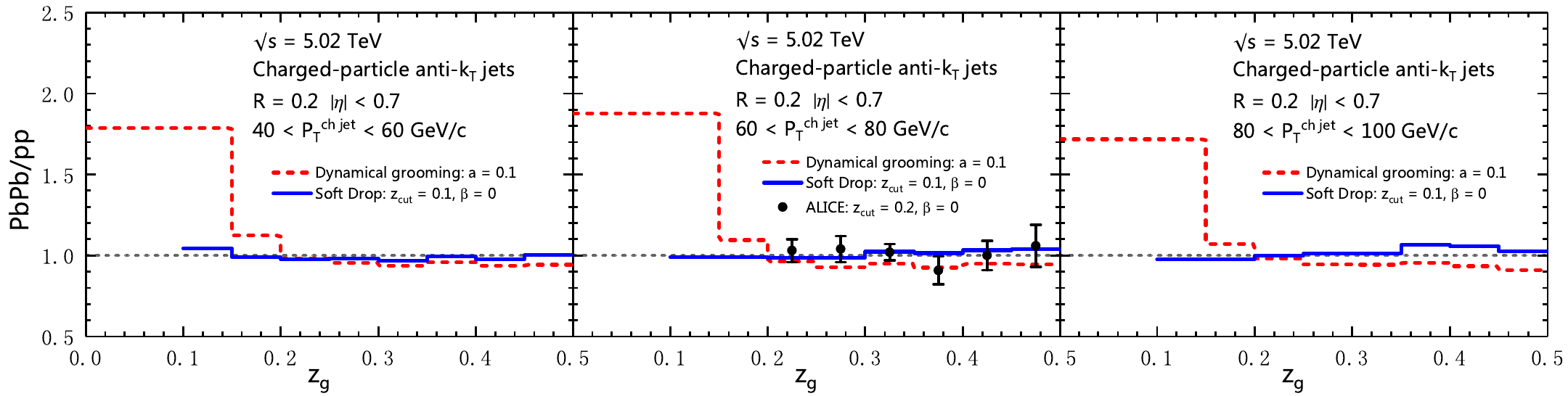}
    \includegraphics[width=0.9\textwidth]{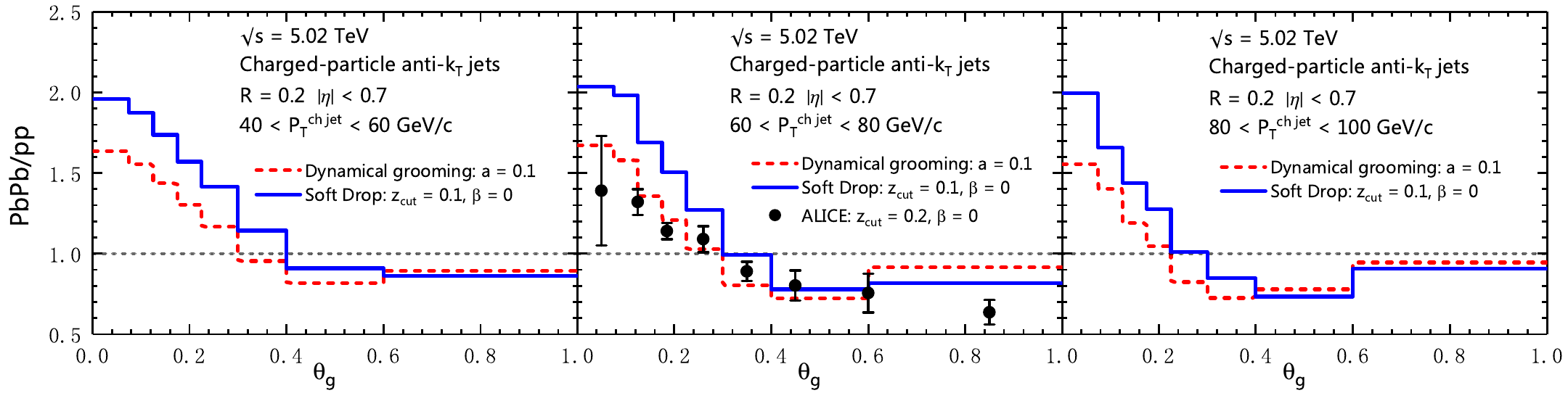}
\caption{The Pb+Pb/p+p ratios of normalized $z_g$ (upper plots) and $\theta_g$ (bottom plots) distribution with the soft drop and the dynamical grooming algorithms at $\sqrt{s} = 5.02$ TeV in different $p_{\mathrm{T,jet}}$ intervals $40 < p_{\mathrm{T,jet}}< 60$ GeV/c, $60 < p_{\mathrm{T,jet}}< 80$ GeV/c and $80 < p_{\mathrm{T,jet}}< 100$ GeV/c from left to right, respectively.}
\label{fig-zgtgraa}
\end{center}
\end{figure*}

 We firstly predict in Fig.~\ref{fig-prediction}, the $z_g$ and $\theta_g$ distributions in central 0-30$\%$ Pb+Pb collisions at $\sqrt{s_{NN}} = 5.02$ TeV for inclusive jets with dynamical grooming algorithms using different dynamical grooming parameter $a=0.1,1,2$. In order to focus on the jet quenching effect, the jet radius parameter $R=0.2$ is selected. In the three cases of groomed parameters $a$, the normalized $z_g$ and $\theta_g$ distributions in Pb+Pb collisions will shift to smaller value compared to that in p+p collisions, leading to enhancement in smaller $z_g$ ($\theta_g$) and suppression in larger $z_g$ ($\theta_g$) as demonstrated in the upper panels. In order to distinguish the distributions for better presentation, different factors has been multiplied to different curves. To better understand the detail of the medium modification, the ratios of the normalized distributions in Pb+Pb and p+p are plotted in the bottom panels. We find, with the decreasing of the parameter $a$, more proportion of the groomed jets in A+A are observed to shift towards smaller $z_g$ compared to that in p+p, and less proportion of the groomed jets in A+A are observed to shift towards smaller $\theta_g$. It indicates that, in the context of dynamical grooming algorithm, the originally triggered more balanced dynamical groomed jets will become more momentum imbalanced and less narrow due to jet quenching.

Next, we compare the medium modification of jet substructures for the groomed jets selected from $a=0.1$ case in the dynamical grooming algorithm and from $z_{\mathrm{cut}}=0.1, \beta =0$ case in the soft drop grooming algorithm scenario. The two cases that we selected to compare can both help generate the most momentum balanced groomed jets in their own categories, and the groomed jets generated from $a=0.1$ case in the dynamical grooming algorithm are much more balanced which is demonstrated in Fig.~\ref{fig-pp}. In Fig.~\ref{fig-zgtgraa}, we plot Pb+Pb/pp ratios of the normalized distributions as functions of $z_g$ (upper panels) and $\theta_g$ (lower panels) in both these two grooming cases of inclusive jets in central 0-30$\%$ Pb+Pb collisions at $\sqrt{s_{\mathrm{NN}}}$ = 5.02~TeV, for three different jet transverse momentum intervals: $\rm 40<p_\mathrm{T}^\mathrm{ch~jet}<60$~GeV, $\rm 60<p_\mathrm{T}^\mathrm{ch~jet}<80$~GeV and  $\rm 80<p_\mathrm{T}^\mathrm{ch~jet}<100$~GeV respectively from left to right panels. In all three different jet transverse momentum intervals, the Pb+Pb/p+p ratios of the normalized $z_g$ distributions within $z_{\mathrm{cut}}=0.1, \beta =0$ in the soft drop grooming algorithm are close to unity, the observation are in consistent with the experimental measurements~\cite{ALargeIonColliderExperiment:2021mqf}. However, within $a=0.1$ in the dynamical grooming case, the Pb+Pb/p+p ratios of the normalized $z_g$ distributions show enhancement in smaller $z_g$ and suppression in larger $z_g$ as demonstrated in the upper plots. It implies that those most balanced jets which are groomed by the dynamical grooming algorithm in p+p are much more modified due to jet quenching compared to that of the soft drop counterparts, leading stronger normalized distribution shifting towards small $z_g$. But the Pb+Pb/p+p ratio comparisons of the two cases with respect to $\theta_g$ in lower plots show that the most balanced jets in p+p would suffer weaker normalized distribution shifting towards small $\theta_g$ due to the medium modification compared to that of the soft drop counterparts. The increasing of jet transverse momentum will lead the stronger distribution shifting of $\theta_g$ towards smaller values. The ALICE experimental data for $z_{\mathrm{cut}}=0.2, \beta =0$ case is also plotted accordingly for reference. To conclude the above observation, the jet quenching effect will lead the groomed jets in both cases become momentum imbalance and size narrower, the originally generated more momentum balanced groomed jets in p+p would become more momentum imbalance and less narrow.

Furthermore, we systematically investigate the jet substructure observables $z_g$ and $\theta_g$ for inclusive jets in p+p and Pb+Pb collisions at $\rm \sqrt{s}=5.02$~TeV in both $z_{\mathrm{cut}} = 0.1$, $\beta = 0$ soft drop grooming case and $a=0.1$ dynamical grooming case. We listed the number of groomed jets averaged momentum sharing of the groomed inclusive jets in p+p and A+A in both two cases for all three jet $\rm P_T$ intervals in Table.~\ref{tab:averzg} and the results of the number of groomed jets averaged groomed jet radius in Table.~\ref{tab:avertg}. We find, in p+p, the averaged splitting fraction $\rm \langle z_g \rangle_{jets}$ (the averaged groomed jet radius $\rm \langle \theta_g \rangle_{jets}$) in $a=0.1$ dynamical grooming case is always larger (smaller) than that in $z_{\mathrm{cut}} = 0.1$, $\beta = 0$ soft drop grooming case in all $p^{\rm ch, jet}_{\rm T}$ intervals, and the higher the $p^{\rm ch, jet}_{\rm T}$ is, the smaller the value of the $\rm \langle z_g \rangle_{jets}$ and the $\rm \langle \theta_g \rangle_{jets}$ are. It means the $a=0.1$ dynamical grooming case always tends to generate more momentum balanced and narrower groomed jets in all $p^{\rm ch, jet}_{\rm T}$ intervals than the $z_{\mathrm{cut}} = 0.1$, $\beta = 0$ soft drop grooming case, and higher $p^{\rm ch, jet}_{\rm T}$ always tends to generate more momentum imbalanced and narrower sized groomed jets in both cases.

The values of the $\rm \langle z_g \rangle_{jets}$ and $\rm \langle \theta_g \rangle_{jets}$ in A+A are consistently smaller than those in p+p for both grooming cases in all three $p^{\rm ch, jet}_{\rm T}$ intervals. It indicates, despite of the different grooming method, the originally generated more balanced groomed jets will become more momentum imbalanced and less jet size narrow due to jet quenching. We find $\rm \langle z_g \rangle^{pp}_{jets}-\langle z_g \rangle^{AA}_{jets} \approx 0.01$ for $a=0.1$ dynamical grooming case, and $\approx 0.02$ for $z_{\mathrm{cut}} = 0.1$, $\beta = 0$ soft drop grooming case in all three $p^{\rm ch, jet}_{\rm T}$ intervals, indicating overall weaker medium modification of $z_g$ for those originally more balanced groomed jets generated in $a=0.1$ dynamical grooming case than that in soft drop case. We conduct the same comparison to $\rm \langle \theta_g \rangle^{pp}_{jets}-\langle \theta_g \rangle^{AA}_{jets}$ for both two cases, find same overall weaker medium modification of $\theta_g$ for those originally more balanced and narrower groomed jets generated in $a=0.1$ dynamical grooming case than that in the soft drop case.

Naively, we expect the soft gluon radiation processes in jet quenching will induce the energy in a jet to distributed at larger radius and the energies of the primary two splitting sub-jets to be more imbalanced. Throughout the investigation, we find the A+A groomed jets always become momentum imbalanced and narrower compared to the p+p counterparts due to quenching for all grooming cases in all three $p^{\rm ch, jet}_{\rm T}$ intervals. It is because of the energy loss effect will induce the jets with originally higher $p^{\rm ch, jet}_{\rm T}$ in p+p which distributed at smaller $z_g$ and $\theta_g$ to fall into the $p^{\rm ch, jet}_{\rm T}$ intervals that we investigated in.   

\begin{table}
    \centering
    \resizebox{0.9\columnwidth}{!}{\renewcommand\arraystretch{1.3}
    \begin{tabular}{|c|c|c|c|}
        \hline
        \multirow{3}{*}{ $P_\mathrm{T}^\mathrm{ch ~ jet}$} & Soft Drop: & Dynamical Grooming:  &  \multirow{3}{*}{} \\
        & $z_{\mathrm{cut}} = 0.1$, $\beta = 0$  & $a=0.1$ &\\
        \cline{2-3}
         &   $\rm \langle z_g \rangle_{jets}$  &   $\rm \langle z_g \rangle_{jets}$    & \\
         \hline
        \multirow{2}{*}{$\rm 40-60$~GeV} & 0.232 & 0.322 & pp \\
        \cline{2-4}
         & 0.210 & 0.312 & AA\\
         \hline
           \multirow{2}{*}{$\rm 60-80$~GeV} & 0.224 & 0.319 & pp \\
        \cline{2-4}
         & 0.200 & 0.308 & AA\\
         \hline
          \multirow{2}{*}{ $\rm 80-100$~GeV} & 0.217 & 0.317 & pp \\
        \cline{2-4}
         & 0.196 & 0.304 & AA\\
         \hline
    \end{tabular}
    }
    \caption{The averaged momentum splitting fraction in inclusive jets with the soft drop grooming and the dynamical grooming algorithms are calculated in both p+p and Pb+Pb collisions  at $\rm \sqrt{s}=5.02$~TeV at three transverse momentum intervals: $\rm 40<p_\mathrm{T}^\mathrm{ch~jet}<60$~GeV, $\rm 60<p_\mathrm{T}^\mathrm{ch~jet}<80$~GeV and  $\rm 80<p_\mathrm{T}^\mathrm{ch~jet}<100$~GeV respectively.}
    \label{tab:averzg}
\end{table}

\begin{table}
    \centering
    \resizebox{0.9\columnwidth}{!}{\renewcommand\arraystretch{1.3}
    \begin{tabular}{|c|c|c|c|}
        \hline
        \multirow{3}{*}{ $P_\mathrm{T}^\mathrm{ch~jet}$} & Soft Drop:  & Dynamical Grooming:  &  \multirow{3}{*}{} \\
        & $z_{\mathrm{cut}} = 0.1$, $\beta = 0$  & $a=0.1$ &\\
        \cline{2-3}
         &   $\rm \langle \theta_g \rangle_{jets}$  &   $\rm \langle \theta_g \rangle_{jets}$    & \\
         \hline
        \multirow{2}{*}{$\rm 40-60$~GeV} & 0.582 & 0.478 & pp \\
        \cline{2-4}
         & 0.539 & 0.443 & AA\\
         \hline
           \multirow{2}{*}{$\rm 60-80$~GeV} & 0.497 & 0.397 & pp \\
        \cline{2-4}
         & 0.441 & 0.358 & AA\\
         \hline
          \multirow{2}{*}{ $\rm 80-100$~GeV} & 0.424 & 0.333 & pp \\
        \cline{2-4}
         & 0.383 & 0.310 & AA\\
         \hline
    \end{tabular}
    }
    \caption{The averaged groomed jet radius in inclusive jets with the soft drop grooming and the dynamical grooming algorithms are calculated in both p+p and Pb+Pb collisions at $\rm \sqrt{s}=5.02$~TeV at three transverse momentum intervals: $\rm 40<p_\mathrm{T}^\mathrm{ch~jet}<60$~GeV, $\rm 60<p_\mathrm{T}^\mathrm{ch~jet}<80$~GeV and  $\rm 80<p_\mathrm{T}^\mathrm{ch~jet}<100$~GeV respectively.}
    \label{tab:avertg}
\end{table}

\section{conclusion}
\label{sum}
In this paper, we systematically predict the normalized distributions of the groomed jet radius $\theta_g$ and momentum splitting fraction $z_g$ with the dynamical grooming algorithms in Pb+Pb collisions at $\sqrt{s_{\mathrm{NN}}}$ = 5.02 TeV respectively. It’s found that in the context of dynamical grooming algorithm, the more balanced dynamical groomed jets generated by smaller parameter $a$ in p+p will become more momentum imbalanced and less jet size narrow due to jet quenching. By comparing these jet substructure observables with different parameters in these two grooming cases in p+p, we find the $a=0.1$ dynamical grooming case would generate the most momentum balanced groomed jets, thus a systematical comparison with the most balanced case in the soft drop grooming $z_{\mathrm{cut}} = 0.1$, $\beta = 0$ is conducted. The normalized distribution ratios Pb+Pb/p+p as respect to $z_g$ in $z_{\mathrm{cut}} = 0.1$, $\beta = 0$ soft drop case are close to unity and those in  $a=0.1$ dynamical grooming case shows enhancement at small $z_g$. Pb+Pb/p+p with respect to $\theta_g$ in the dynamical grooming case demonstrate weaker modification than those in the soft drop counterparts. We further calculate the groomed jet number averaged momentum splitting fraction $\rm \langle z_g \rangle_{jets}$ and averaged groomed jet radius $\rm \langle \theta_g \rangle_{jets}$ in p+p and A+A for both grooming cases in all three $p^{\rm ch, jet}_{\rm T}$ intervals. We find the originally generated more balanced groomed jets will become more momentum imbalanced and less jet size narrow due to jet quenching, and weaker medium modification of $z_g$ and $\theta_g$ in $a =0.1$ dynamical grooming case than in the soft drop counterpart. 

\textbf{Acknowledgments:} This research is supported by the Guangdong Major Project of Basic and Applied Basic Research No. 2020B0301030008, the Natural Science Foundation of China with Project Nos. 11935007 and 11805167. 

	%
\end{document}